\documentstyle[11pt,IAUS215,twoside,epsf]{article}

\def\edcomment#1{\iffalse\marginpar{\raggedright\sl#1\/}\else\relax\fi}
\reversemarginpar

\begin{document}
\vspace*{1cm}
\title{Evolution of massive stars with rotation and mass loss}
 \author{Andr\'e Maeder  \& Georges Meynet}
\affil{Geneva Observatory, CH--1290 Sauverny, Switzerland}

\begin{abstract}
Rotation appears as a dominant effect in massive star evolution. It
largely affects all the model outputs: inner structure, tracks, lifetimes, isochrones,
surface compositions, blue to red supergiant ratios, etc. At lower metallicities,
the effects of rotational mixing are larger; also, more stars
may reach critical velocity, even if 
the initial  distribution of rotational velocities is the same.
\end{abstract}

\section{Introduction}
In this review, we shall concentrate mostly  on the rotational  properties of massive
stars.  Rotation has an
importance which for massive stars in the Galaxy is comparable to that of mass loss by stellar winds.
For massive stars at lower metallicities, it is likely that rotation plays an  even
more important role.

\section{Structure, mixing and mass loss of rotating stars}

\subsection{Structure}
The equations of stellar structure for a rotating star were written by 
Kippenhahn \&  Thomas (1970). These equations  apply to a star in solid body or cylindrical 
rotation. For stars in differential rotation, the use of the above equations is incorrect.
 Here, we consider
the case of shellular rotation with $\Omega=\Omega(r)$ (Zahn 1992). In this case, the 
structure equations need  to be   modified (Meynet \& Maeder 1997). 
Structural effects due to the centrifugal force
are in general small in the interior. However, the  distorsion of stellar surface may be large
enough to produce significant shifts in the HR diagram  (Maeder \& Peytremann 1970).

\subsection{Internal transport}
For the transport of chemical elements, we consider the effects of
shear mixing, of meridional circulation and their interactions  
with the  horizontal turbulence.  For shellular rotation, the equation of transport of angular
momentum in the vertical direction is in lagrangian coordinates
(Zahn 1992),
\begin{eqnarray}
\rho \frac{d}{d t}
\left( r^2 \Omega\right)_{M_r} = 
 \frac{1}{5 r^2}  \frac{\partial}{\partial r}
\left(\rho r^4 \Omega \; U(r) \right)
  + \frac{1}{r^2} \frac{\partial}{\partial r}
\left(\rho D_{\mathrm{shear}}\; r^4 \frac{\partial \Omega}{\partial r} \right) .
\end{eqnarray} 

\noindent $\Omega(r)$ is the mean angular velocity at level $r$. $U(r)$ is the
 radial term of the vertical component of the velocity of the meridional
circulation. $D_{\mathrm{shear}}$ is the coefficient of shear diffusion. From the two terms in the 
second member, we see that {\emph{advection and diffusion are not the same}. Thus, 
the transport of angular momentum by circulation cannot be 
treated as a diffusion.
For the changes of the chemical 
elements due to transport, we may use a diffusion equation  with a diffusion coefficient 
which is  the sum  of  $D_{\mathrm{shear}}$, as given below, 
and  of $D_{\rm eff} = \frac{\mid rU(r) \mid^2}{30 D_h}$. $D_{\rm eff}$ expresses the resulting
effect of meridional circulation and of a large horizontal turbulence
(Chaboyer \& Zahn 1992). This expression of $D_{\rm eff}$ tells us that the vertical
advection of chemical elements is  inhibited by the
strong horizontal turbulence characterized by $D_{\rm{h}}$. 
 The usual estimate of $D_{\mathrm{h}} =
\frac{1}{c_{\mathrm{h}}} r \;|2V(r) - \alpha U(r)| $ was given by 
Zahn (1992). A recent study suggests that this coefficient 
is at least an order of magnitude larger (Maeder 2002),
\begin{eqnarray}
D_{\mathrm{h}} =  A \; r \; \left(r
\overline{\Omega}(r) \; V \;
 \left[ 2 V - \alpha U \right]\right)^\frac{1}{3}  
\quad  \mathrm{with} \quad   A = \left( \frac{3}{400 n \pi} \right)^{\frac{1}{3}} \;.
\label{nuh}
\end{eqnarray}

\noindent
For n=1, 3 or 5 $A \approx 0.134, 0.0927, 0.0782$ respectively.
For the coefficient $D_{\mathrm{shear}}$, we use the results by Maeder (1997), Talon \& Zahn (1997)
and   Maeder \& Meynet (2001). $D_{\mathrm{shear}}$
is also modified by the horizontal turbulence  $D_{\mathrm{h}}$, in the sense that a larger
$D_{\mathrm{h}}$
 leads to a  decrease of the mixing of elements, 
\begin{eqnarray}
D_{\mathrm{shear}} =  \frac{ (K + D_{\mathrm{h}})}
{\left[\frac{\varphi}{\delta} 
\nabla_{\mu}(1+\frac{K}{D_{\mathrm{h}}})+ (\nabla_{\mathrm{ad}}
-\nabla_{\mathrm{rad}}) \right] }\; \times 
 \frac{H_{\mathrm{p}}}{g \delta} \; 
\left [ \alpha\left( b \; \Omega{d\ln \Omega \over d\ln r} \right)^2
-4 (\nabla^{\prime}  -\nabla) \right] \; ,
\end{eqnarray}

\noindent
with b=0.8836. The radial component $U(r)$ in Eqs. (1) and (2) was given by Maeder \& Zahn (1998),
\begin{eqnarray}
U(r) &=&  \frac{P}{\overline{\rho} \overline{g} C_{\!P} \overline{T}
\, (\nabla_{\rm ad} - \nabla +  (\varphi/\delta) \nabla_{\mu}) }
\times   {  \frac{L}{M_\star} \left[E_\Omega + E_\mu \right] + \!
\frac{C_P}{\delta} \frac{\partial \Theta }{\partial t} \! } ,
\end{eqnarray}

\noindent
where $M_{\star}=M \left( 1 - \frac{\Omega^2}{2 \pi G \rho_{\rm{m}}}  \right)$
is the reduced mass,
with the  notations  given in the quoted paper.
The driving term in the square brackets in the second member
 is $E_{\Omega}$.  It behaves mainly like,
\begin{equation}
E_{\Omega}  \simeq  \frac{8}{3} \left[ 1 - \frac{{\Omega^2}}
{2\pi G\overline{\rho}}\right] \left( \frac{\Omega^2r^3}{GM}
\right) .
 \end{equation}

\noindent
The term $\overline{\rho}$ means the average
on the considered equipotential.
The term with the minus sign in the square bracket is the 
Gratton--\"{O}pik term, which becomes important in the outer layers
due to the decrease of the local density. It
can produce negative values of $U(r)$. A  negative $U(r)$
means a circulation going down along the polar axis and up 
in the equatorial plane. This makes an outwards transport 
of angular momentum, while a positive $U(r)$ gives an inward transport
of angular momentum. Before,
 the term $\nabla_{\mu}$ was never
properly included. The justification for this term is not so 
straightforward (Maeder \& Zahn 1998).
\begin{figure}
\plottwo{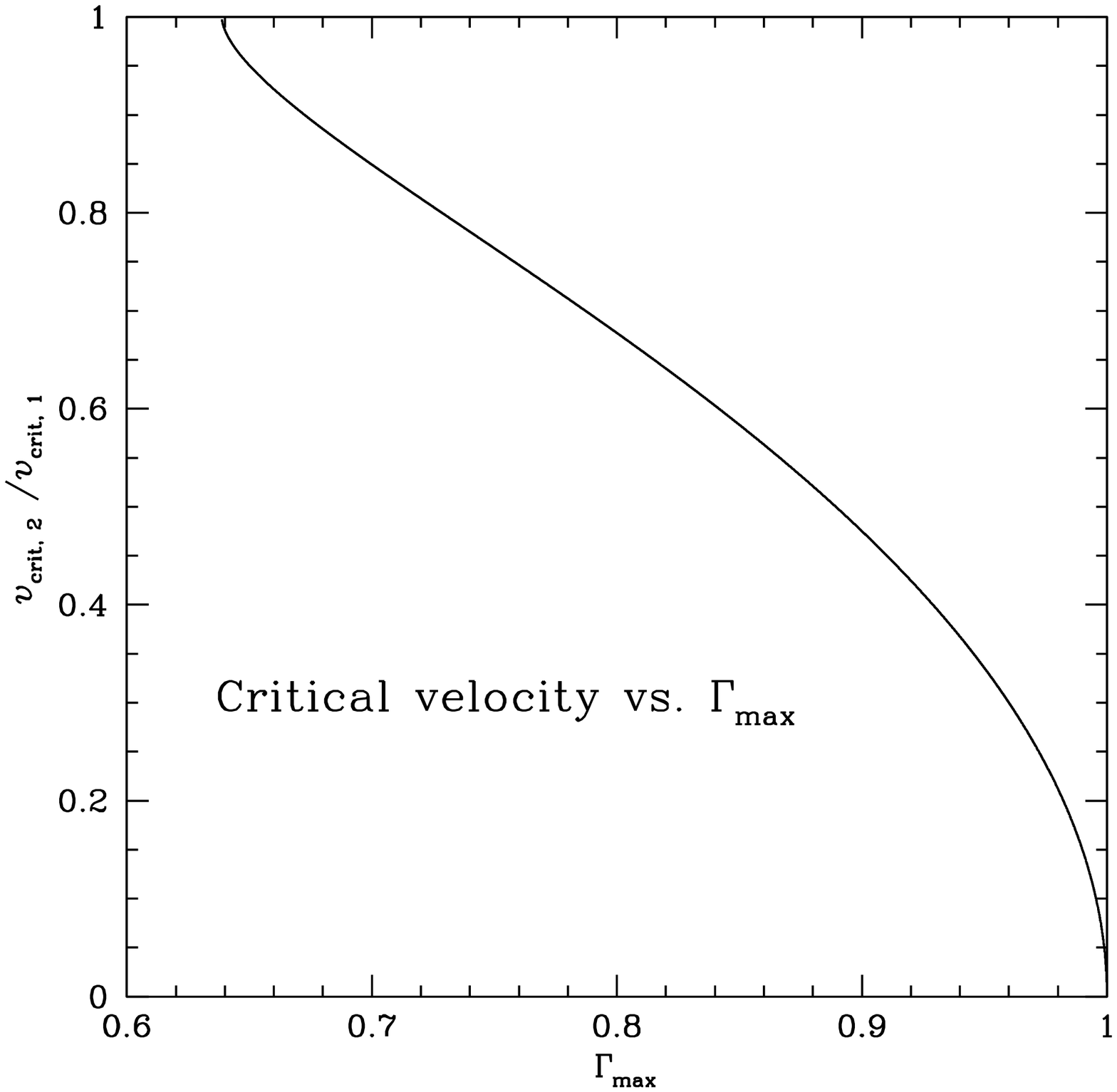}{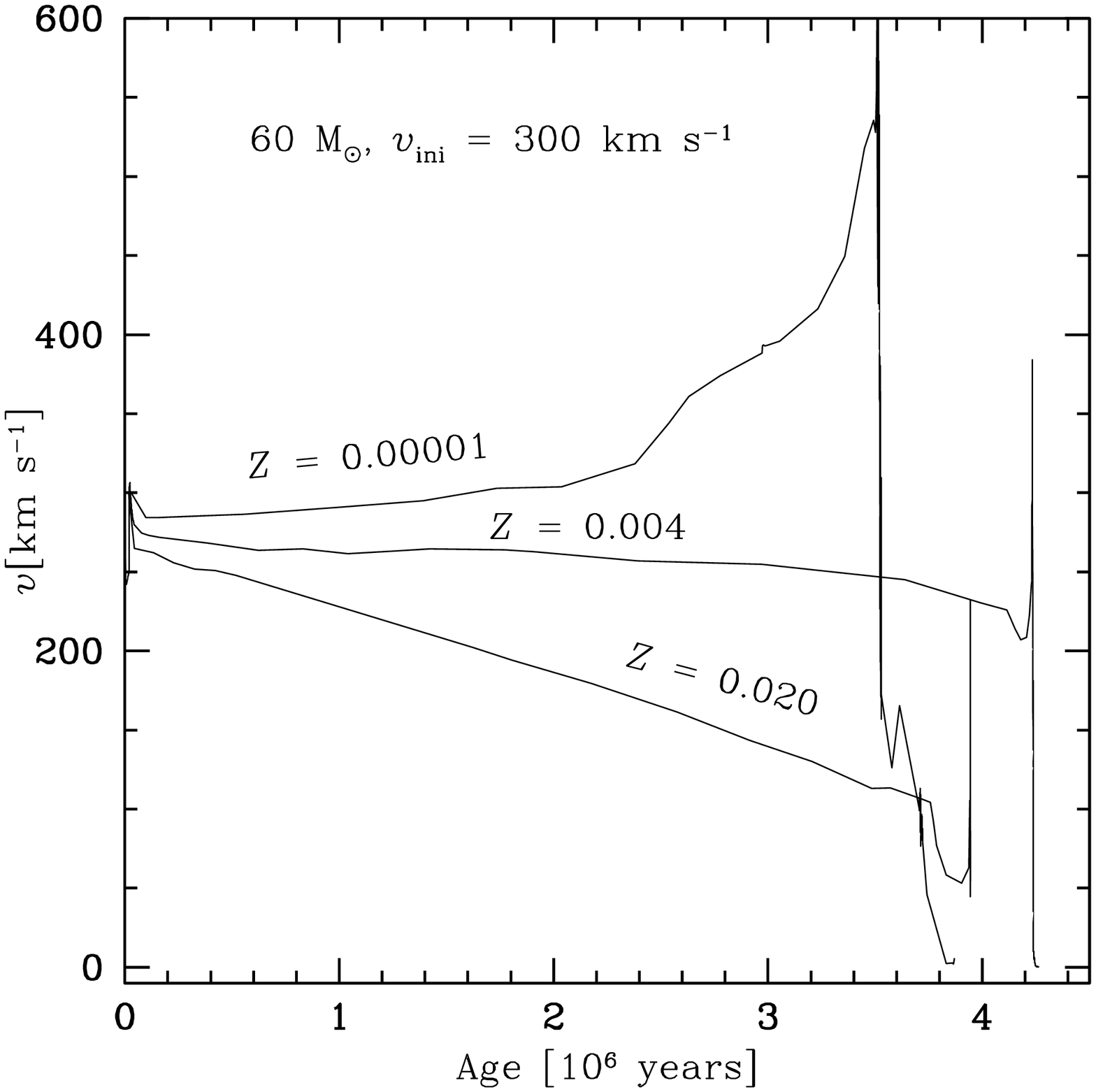}
\caption{Left: The second critical velocity $v_{\mathrm{crit, 2}}$ expressed as
a fraction of $v_{\mathrm{crit, 1}}$ plotted as a function
of the Eddington factor   $\Gamma_{\mathrm{max}}$, which is 
the largest value of the Eddington factor over the stellar surface. 
 Right: Evolution of the surface velocities for a 60 M$_
 {\odot}$ star with 3 different
 initial metallicities.}
\end{figure}

\subsection{Critical velocity and mass loss in a rotating star}

For a rotating star, one must consider 
the flux $F(\vartheta)$ at a given colatitude $\vartheta$ as given by von Zeipel's theorem.
This theorem  has been studied for a star with shellular rotation by 
Maeder (1999), who finds $F(\vartheta)  =  - \frac{L(P)}{4 \pi GM_{\star}}
g_{\rm{eff}} [1 + \zeta(\vartheta)]$,  which shows a small extra--term
 $\zeta(\vartheta)$ depending on $\Omega, T, \nabla T, \rho$ and opacities.
 In a rotating star, the Eddington factor becomes a local quantity $\Gamma_{\Omega}(\vartheta)$.
 We define it  as
the ratio of the local flux $F(\vartheta)$ 
given by the von Zeipel theorem to the maximum possible local flux,
which is $F_{\mathrm{lim}}(\vartheta) = - \frac{c}{\kappa(\vartheta)}  
g_{\mathrm{eff}}(\vartheta)$. Thus, one has
\begin{eqnarray}
\Gamma_{\Omega}(\vartheta) =
\frac{F(\vartheta)}{F_{\mathrm{lim}}(\vartheta)}=
\frac{ \kappa (\vartheta) \; L(P)[1+\zeta (\vartheta)]}{4 \pi 
cGM \left( 1 - \frac{\Omega^2}{2 \pi G \rho_{\rm{m}}}  \right) } \; ,
\end{eqnarray}

\noindent 
where the opacity $\kappa(\vartheta)$ depends on the colatitude $\vartheta$, since $T_{\mathrm{eff}}$
also depends $\vartheta$.
The Eddington factor  depends on
the angular velocity $\Omega$ on the isobaric surface. 
This shows that the maximum luminosity of a rotating star is decreased by rotation.
It is to be stressed that if the limit  $\Gamma_{\Omega}(\vartheta) = 1 $ 
happens to be met in general at the equator, it is not because 
$g_{\mathrm{eff}}$ is the lowest there, but because the 
opacity is the highest! Indeed, the  dependences 
of $F(\vartheta)$ and of $F_{\mathrm{lim}}(\vartheta)$ with respect to $g_{\mathrm{eff}}$
have cancelled each other in $ \Gamma_{\Omega}(\vartheta)$.

Often, the critical velocity in a rotating star is written as
$v^2_{\rm{crit}} = \frac{GM}{R} (1-\Gamma)$. This expression is incorrect,
since it would apply only to uniformly bright stars. 
The critical velocity of a rotating star is given by the zero of the equation
expressing the total gravity 
$\vec{g_{\mathrm{tot}}} = \vec{g_{\mathrm{eff}}} + \vec{g_{\mathrm{rad}}} =
 \vec{g_{\mathrm{grav}}} + \vec{g_{\mathrm{rot}}} + \vec{g_{\mathrm{rad}}}$.
 This is 
 \begin{equation}
\vec{g_{\mathrm{tot}}} = \vec{g_{\mathrm{eff}}}
\left[ 1 - \Gamma_{\Omega}(\vartheta) \right] \; .
\end{equation}

\noindent 
This equation has two roots (Maeder \& Meynet 2000). The first that is met determines the critical 
velocity. The first root is as usual  $v_{\mathrm{crit, 1}} = 
\left( \frac{2}{3} \frac{GM}{R_{\mathrm{pb}}} \right)^{\frac{1}{2}}$, where
$R_{\mathrm{pb}}$ is the polar radius at break--up.
The second root applies to Eddington factors bigger than 0.639. It
is illustrated in Fig. 1 (left), which  shows that, when $\Gamma$ is close to
1.0, the critical velocity goes to zero.
\begin{figure}
\plottwo{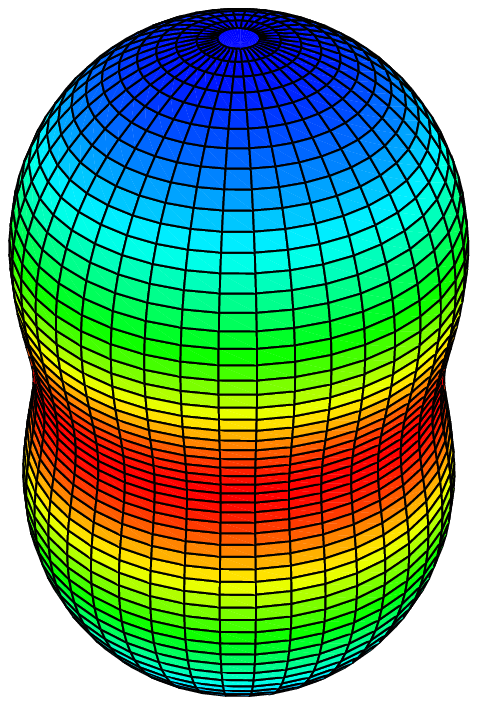}{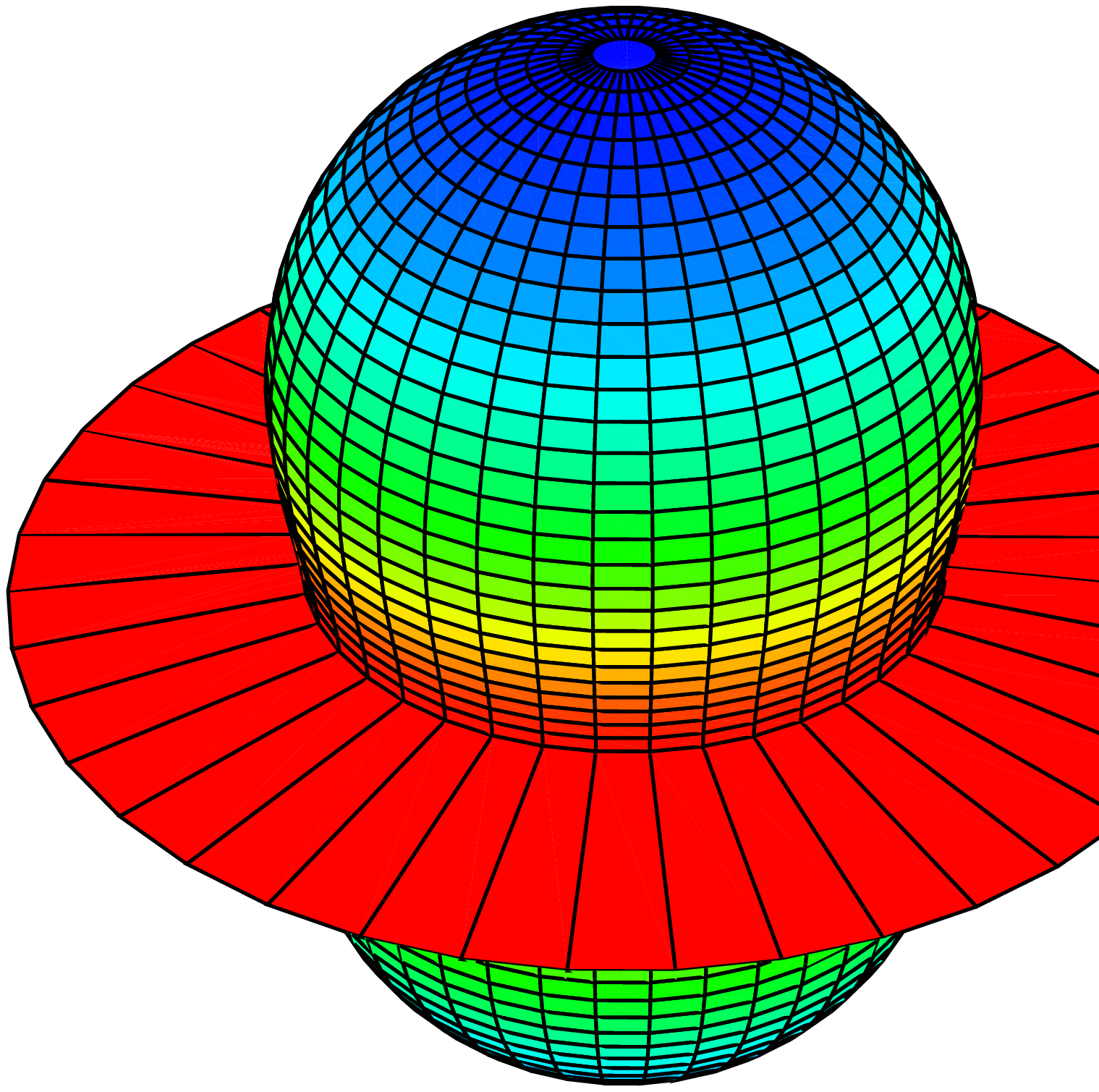}
\caption{Left: Iso-mass loss distribution for a 120 M$_{\odot}$ 
star with $\log{\frac{L}{L_{\odot}}=6.0}$
and  T$_{\mathrm{eff}}=30 000$ K rotating at a fraction 0.80 of break--up velocity. Right:
The same  with T$_{\mathrm{eff}}=25 000$ K, (Maeder \& Desjacques 2000).}
\end{figure}
The theory of radiative winds applied to a  rotating star 
leads to an expression (Maeder 1999) of the mass flux as a function of 
colatitude. Figs. 2 illustrate the distribution of the mass loss rates
around a massive star of 120 M$_{\odot}$ with two different $T_{\mathrm{eff}}$.
For a star hot enough to have electron scattering opacity as the dominant opacity
source from pole to equator, the iso--mass loss curve has a peanut--like 
shape (Fig. 2 left). This  results from the fact that the pole is hotter 
(``$g_{\mathrm{eff}}$--effect'').
If the  $T_{\mathrm{eff}}$ of the star is lower,  a bistability limit
(i.e. a steep increase of the opacity, cf. Lamers 1995) may occur somewhere between the 
pole and the equator, due to the decrase of $T_{\mathrm{eff}}$ from pole
to equator. This ``opacity--effect'' produces an equatorial enhancement
of the mass loss (Fig. 2 right). The anisotropies of mass loss influence the loss of angular
momentum, in particular polar mass loss removes mass but relatively little 
angular momentum. This may strongly influence the evolution (Maeder 2002; see Fig. 5 left below).
We may estimate the mass loss rates of a rotating star 
compared to that of a non--rotating star at the same location in the HR
diagram. The result is (Maeder \& Meynet 2000)
\begin{equation}
\frac{\dot{M} (\Omega)} {\dot{M} (0)} \simeq
\frac{\left( 1  -\Gamma\right)
^{\frac{1}{\alpha} - 1}}
{\left[ 1 - 
\frac{4}{9} (\frac{v}{v_{\mathrm{crit, 1}}})^2-\Gamma \right]
^{\frac{1}{\alpha} - 1}} \; ,
\end{equation}

\noindent
where $\Gamma$ is the electron scattering opacity for a non--rotating
star with the same mass and luminosity, $\alpha$ is a force multiplier (Lamers et al. 1995). Values
of $\alpha$ are given in Table 1,
which shows, for stars of different 
masses and $\Gamma$, the values of the ratio of the mass loss
rates for a  star at critical rotation to that of a non--rotating star of the same M, L and $T_{\mathrm{eff}}$.
\begin{figure}
\plottwo{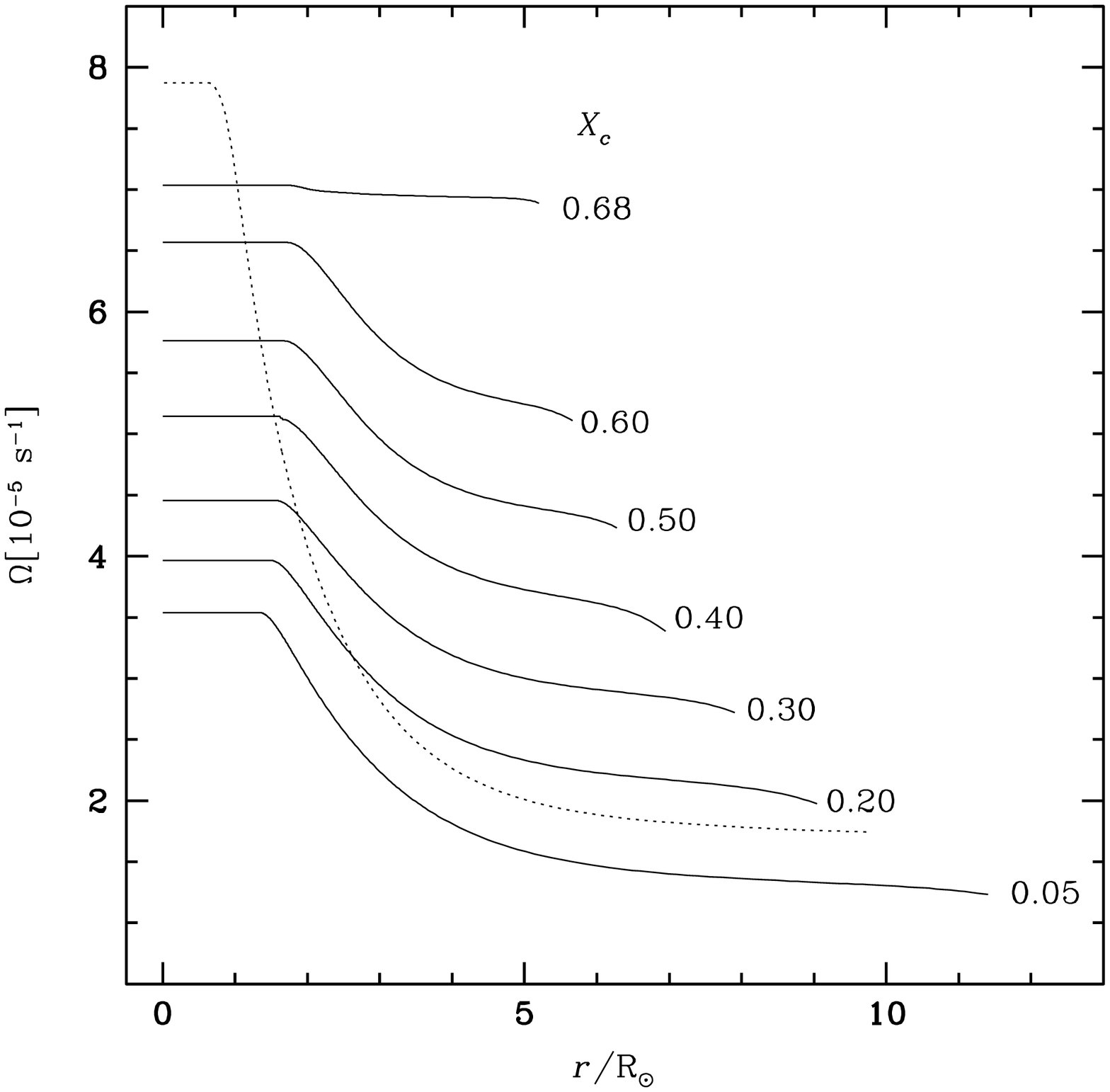}{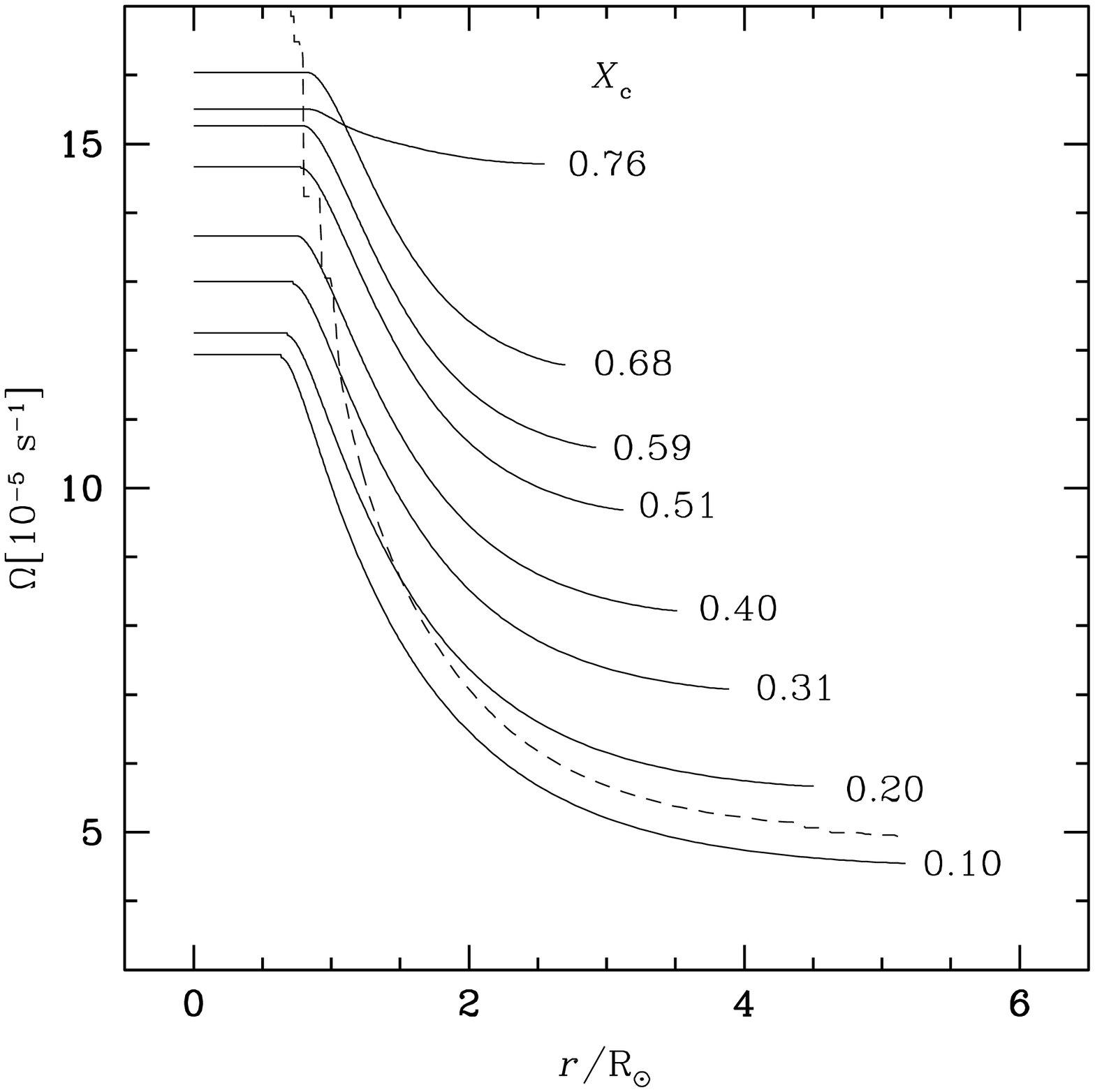}
\caption{Left: Evolution during the MS phase
of the angular velocity in the interior of a 20 M$_{\odot}$
star at $Z$=0.02. Right: the same for a 15 M$_{\odot}$ star at  $Z=10^{-5}$. The values of
the central H--content $X_{\mathrm{c}}$ are indicated.
(From Meynet \& Maeder 2000, 2002).}
\end{figure}
The values $\infty$ indicate that the critical velocity $v_{\mathrm{crit,2}}$ is met.
Close to
$\Gamma=1$ the  increase of the mass loss rates may be quite large. For
blue and red supergiants, the increase could also be large, however supergiants
do not rotate very fast in general. 

We must distinguish 3 cases of stellar break--up: 1.-- The
$\Gamma$--Limit, when radiation effects largely dominate; 2.-- The $\Omega$--Limit, when
rotation effects are essentially determining break--up  and 3.-- The $\Omega \Gamma$--Limit, when both
rotation and radiation are important for the critical velocity and Eq. (8) applies. For
very  massive stars, the  $\Omega \Gamma$--Limit
is clearly the relevant case.
\begin{table}
\caption{Values
of $\frac{\dot{M} (\Omega)} {\dot{M} (0)}$ for different stellar masses. 
The four force multipliers $\alpha$ apply respectively to stars with:
$ 4.70 \geq \log T_{\mathrm{eff}} \geq 4.35$ (type B1.5
or earlier),  $\log T_{\mathrm{eff}}$ = 4.30 (type B2.5),
 4.00 (B9.5), 3.90 (A7) respectively.}
\begin{center}\scriptsize
\begin{tabular}{ccccccc}
$M_{\rm ini}$ &  $\Gamma$
&    $\frac{\dot{M} (\Omega)} {\dot{M} (0)}$
&    $\frac{\dot{M} (\Omega)} {\dot{M} (0)}$
&    $\frac{\dot{M} (\Omega)} {\dot{M} (0)}$

&    $\frac{\dot{M} (\Omega)} {\dot{M} (0)}$ \\


 &  & $\alpha$ = 0.52 & $\alpha$ = 0.24 & $\alpha$ = 0.17
 & $\alpha$ = 0.15 \\
\hline
      &      &         &         &       \\
 120 & 0.903 & $\infty $ & $\infty$  & $\infty $ & $\infty $  \\
 85  & 0.691 & $\infty $ & $\infty$  & $\infty $ & $\infty $  \\

  60 & 0.527 & 3.78  & 96.2  &  1130  & 3526  \\
  40 & 0.356 & 2.14  & 13.6  &  55.3  & 106.0 \\
  25 & 0.214 & 1.76  &  7.02 &  20.1  &  32.6 \\
  20 & 0.156 & 1.67  &  5.87 &  15.2  &  23.6 \\
  15 & 0.097 & 1.60  &  5.04 &  12.1  &  18.1 \\
  12 & 0.063 & 1.57  &  4.68 &  10.8  &  15.8 \\
   9 & 0.034 & 1.54  &  4.41 &   9.8  &  14.2 \\
     &       &       &       &       \\

\hline
\end{tabular}
\end{center}
\end{table}

\section{Evolution of rotational velocities}

Fig. 3  shows the evolution during the MS phase of the internal 
angular velocities in  models at $Z=0.02$ (left)
and at $Z=10^{-5}$ (right). We see that the decrease of the core rotation
is stronger at higher $Z$. This is due to the larger mass loss which removes
more angular momentum, (the case of isotropic mass loss is considered here with Eq. (8)
above). We note that the gradient of $\Omega$ built during MS evolution outside the core 
is larger at lower $Z$. There are 2 reasons for that. One is the higher compactness of the 
star at lower $Z$. The second one is more subtle.  At $Z=0.02$, the density of the outer layers
is lower and the star is more extended than at  lower $Z$, thus the Gratton--\"{O}pik term
is more important. This produces an outward transport of angular momentum and smoothens
the $\Omega$--gradient at $Z=0.02$ more than at lower $Z$. The steeper $\Omega$--gradient 
at lower $Z$ is leading to stronger mixing of chemical elements, as shown below.
\begin{figure}
\plotfiddle{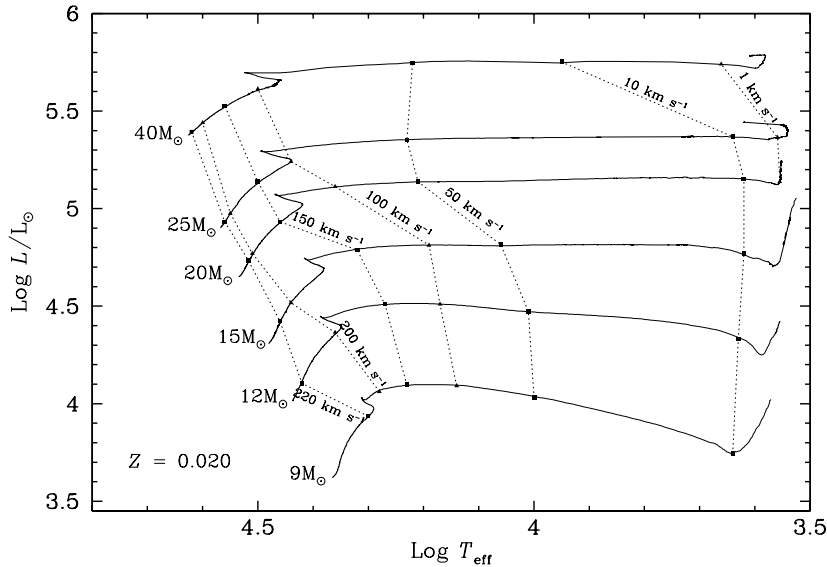}{7cm}{-90}{40}{40}{-170cm}{230cm}
\caption{Evolution  of the equatorial velocities along the redward parts of the tracks
in the HR diagram for initial velocities of 300 km s$^{-1}$ and $Z=0.02$, (Meynet \& Maeder 2000).}
\end{figure}

The evolution of the surface velocities results from internal transport 
and surface losses of angular momentum. Fig. 1 (right) shows the evolution
of the surface velocities for a 60 M$_{\odot}$ model for 3 very different
metallicities $Z$. For such a large mass, the high  losses of mass and angular momentum
lead to a strong decrease of  rotation at $Z=0.02$. On the contrary, at lower 
$Z$ the velocities are growing, as a result of the  contraction of the
stellar core and of the relatively good coupling of the outer layers
produced  by circulation. Thus, it is likely that {\emph{at low $Z$, 
rotation may be a dominant effect
in massive star evolution}}. These stars may reach break--up and then loose
 a significant amount of matter.
Paradoxically, low $Z$ stars may loose mass due to fast rotation as a consequence 
of their initial slow winds!
\begin{figure}
\plottwo{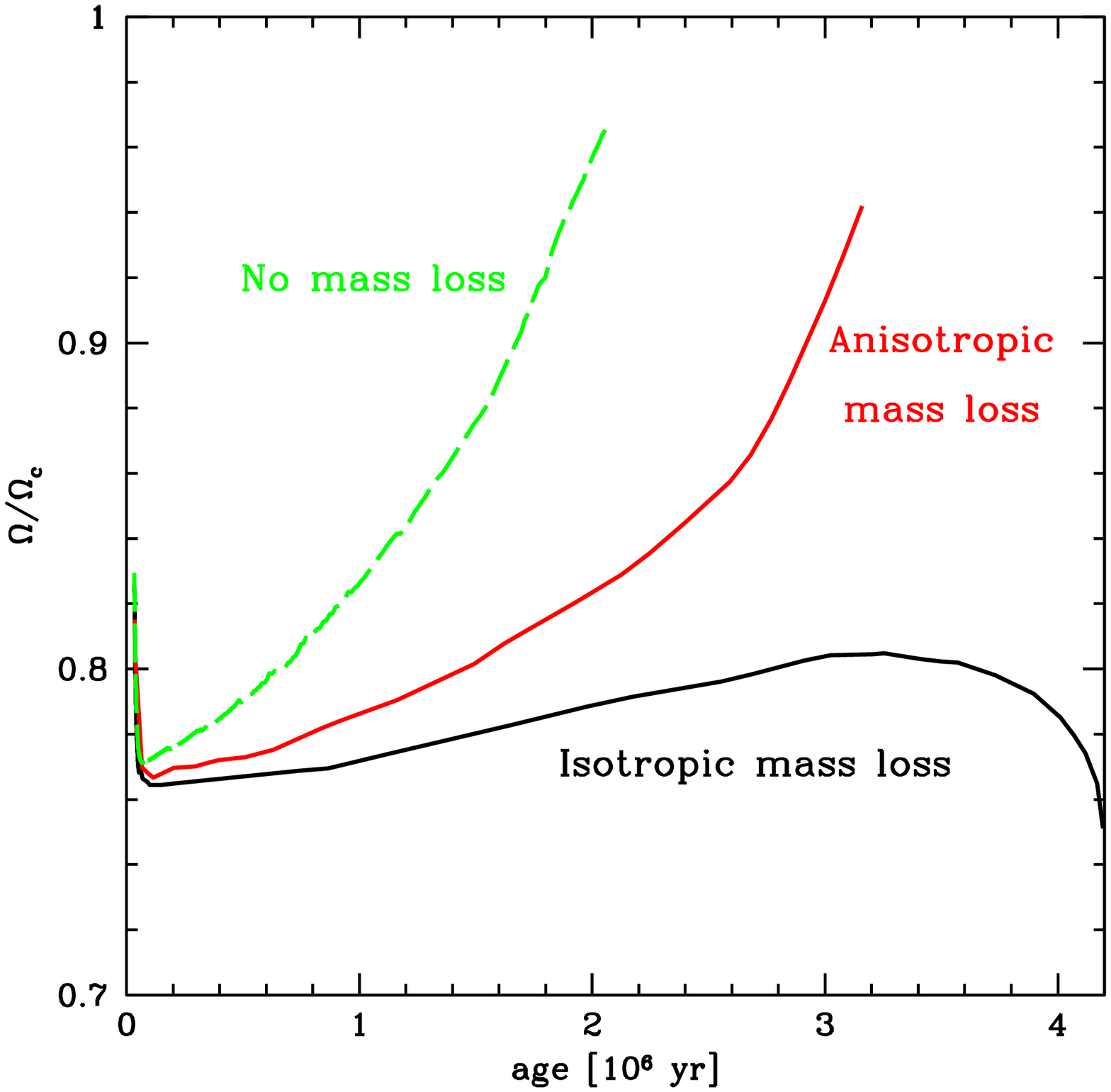}{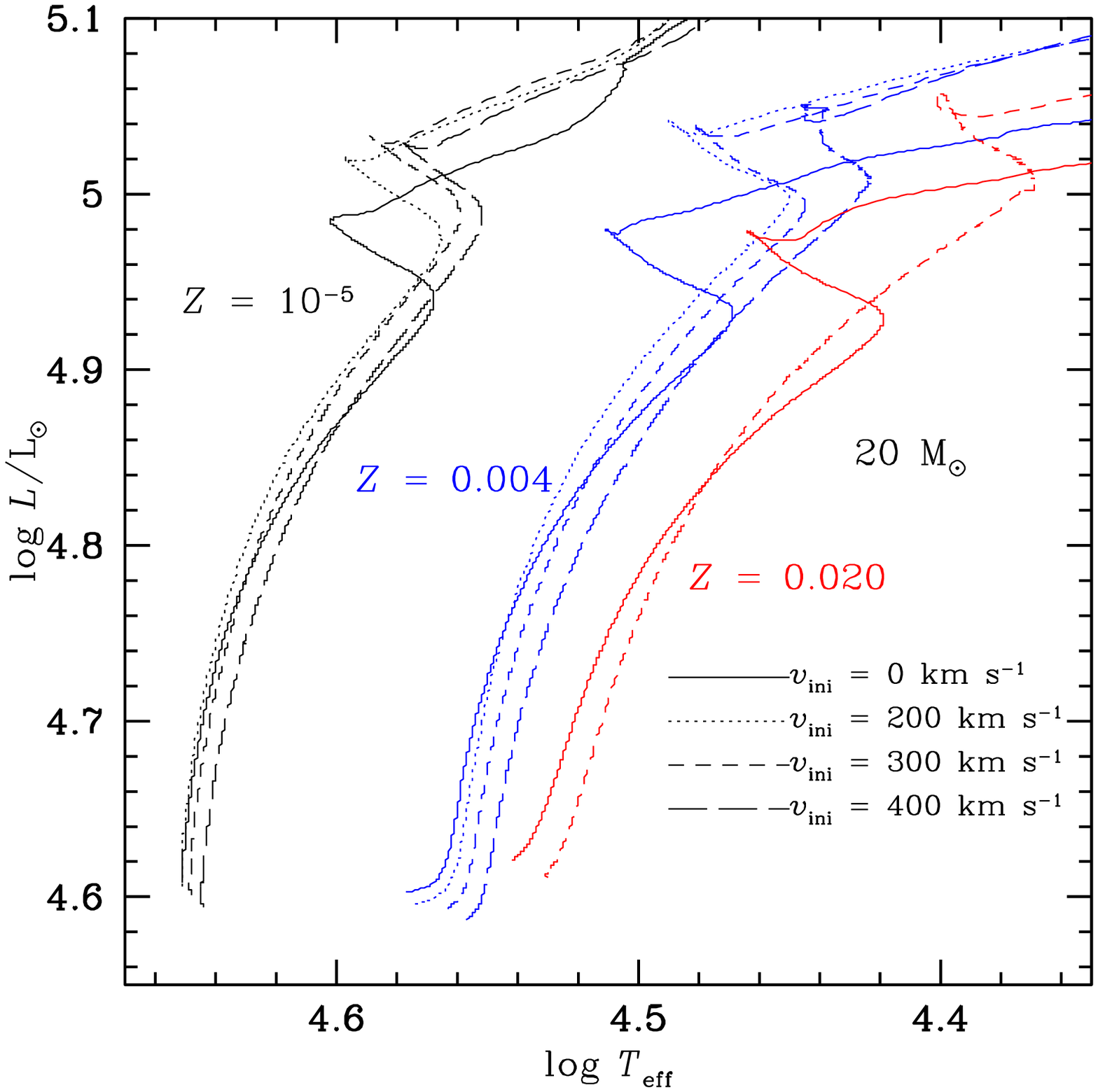}
\caption{Left: Evolution of the ratio $\frac{\Omega}{\Omega_{\mathrm{c}}}$ of the angular velocity to the critical value during the MS phase of a 40 M$_{\odot}$ star with an initial velocity of 500 km s$^{-1}$
(Maeder 2002).
 Right: Evolutionary tracks on the MS for 20 M$_{\odot}$ models of
 different $Z$ and initial velocities $v_{\mathrm{ini}}$, (Meynet \& Maeder 2002).}
\end{figure}

In addition, we may wonder whether the initial distribution of stellar rotation
velocities is the same at lower $Z$. There are some reasons (Maeder et al. 1999) to suspect that 
the initial rotation velocities may be faster at lower $Z$.

Fig. 4 shows the lines of iso--rotational velocities in the HR diagram 
of massive stars at $Z=0.02$ for an initial velocity of 300 km/s. We see
an important decrease for the most massive stars. 
This provides the basis for interesting comparison tests.
 Let us emphasize that the decline will be less important when the anisotropic losses of angular
momentum are accounted for, which is not the case in Fig. 4. Fig. 5 (left) 
shows the very different evolution of the surface rotation of fast rotating stars for different
cases of mass loss. For zero--mass loss, critical velocities are reached very quickly.
For isotropic mass loss (here the relatively lower rates by Vink et al. 2001), there is no increase
of $\frac{\Omega}{\Omega_{\mathrm{c}}}$.
The case  with account of the anisotropic losses of angular momentum gives results 
intermediate between the two previous cases, with some significant increase of 
$\frac{\Omega}{\Omega_{\mathrm{c}}}$.
If there is some magnetic coupling,  
$\frac{\Omega}{\Omega_{\mathrm{c}}}$ will be larger during evolution. Thus, the
study of rotational velocities for stars at various distances from the ZAMS may provide a powerful
test on the role of magnetic field.
\begin{figure}
\plottwo{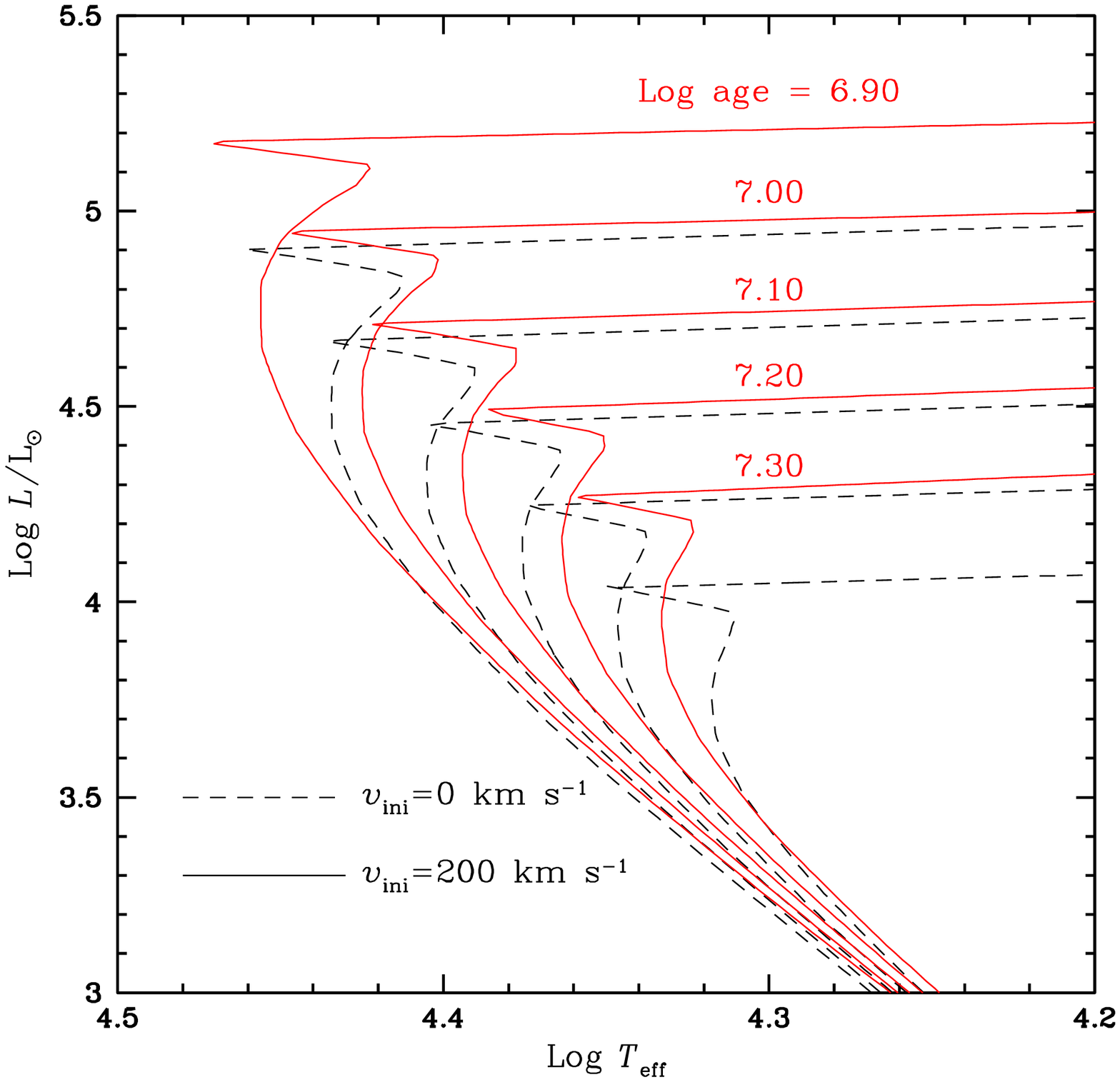}{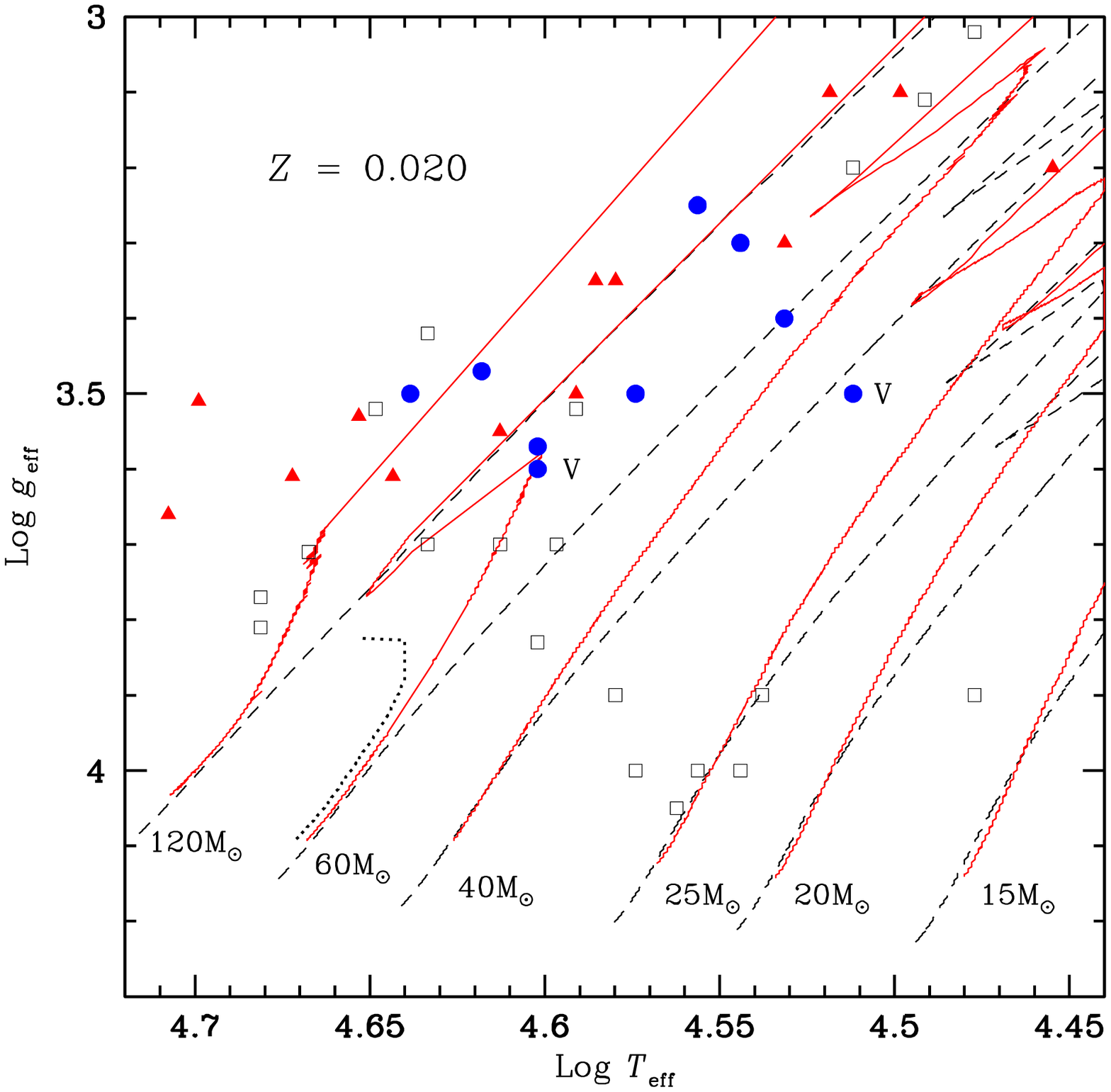}
\caption{Left: Isochrones with and without rotation for solar metallicity.  
 Right: Evolutionary tracks in the log g$_{\mathrm{eff}}$  vs.
 log T$_{\mathrm{eff}}$ diagram. The dashed lines correspond to non--rotating
 tracks, the continuous lines are for models with an initial rotation of 300 km s$^{-1}$.
 The small dotted line corresponds to a very fast rotating 60 M$_{\odot}$.
 The observations are by Herrero et al. (2000) and previous works.}
\end{figure}

\section{HR diagram, lifetimes and isochrones}

Table 2 provides various information on the effects of rotation for a 20 M$_{\odot}$
star. We show the average velocities during the MS phase, the MS lifetimes, the masses 
and velocities at the end of the MS, the helium content, the N/C and N/O ratios at the end of the MS. 
We notice   the growth of the MS lifetimes for higher rotation velocities, as well as the lower final
masses. This results from both the longer MS lifetimes and from the higher mass loss rates.
We notice the chemical enrichments at the stellar surface. 
Some isochrones calculated for initial velocities $v_{\mathrm{ini}}=200$ km/s are 
displayed in Fig. 6 (left); these $v_{\mathrm{ini}}$ lead to average velocities on the 
MS of about 140 km/s.  If we assign cluster ages from these isochrones,
 we obtain ages typically 25\% larger than from the
standard models without rotation. For average velocities of about 220 km/s,
the difference on the age estimate would  be  larger.
This effect  may easily help us to reconcile the ages 
for the Pleiades determined from the turnoff with the ages from lithium
abundances in the low mass stars of the Pleiades  (Martin et al. 1998).
\begin{table}
\caption{Properties of 20 M$_\odot$ models at the end of the MS for different initial
velocities. The velocities are in km s$^{-1}$, the lifetimes in million years,
the masses in solar mass and the surface abundances in mass fraction.} 
\begin{center}\scriptsize
\begin{tabular}{ccrccccc}
$v_{\rm ini}$ & $\overline{v}$ & $t_H$ & M & $v$ & He & N/C & N/O \\
      &      &         &         &       &       &       &      \\
\hline
      &      &         &         &       &       &       &      \\
 0 &       0    & 7.350  & 19.019 & 0     & 0.30  & 0.25  & 0.12 \\
 50 &      30    & 7.720  & 18.896 & 18    & 0.30  & 0.27  & 0.12 \\
 100 &      62    & 8.292  & 18.681 & 46    & 0.30  & 0.45  & 0.19 \\
 200 &     132    & 8.901  & 18.324 & 94    & 0.32  & 1.01  & 0.38 \\
 300 &     197    & 9.309  & 18.020 & 167   & 0.35  & 1.77  & 0.58 \\
 400 &     253    & 9.745  & 17.646 & 217   & 0.37  & 2.54  & 0.76 \\
 500 &     294    & 10.275 & 17.181 & 213   & 0.40  & 3.65  & 0.99 \\
 580 &     304    & 10.324 & 17.148 & 214   & 0.39  & 3.75  & 1.00 \\
     &       &         &         &       &       &       &      \\

\hline
\end{tabular}
\end{center}
\end{table}

Fig. 6 (right) shows the tracks in the $\log g_{\mathrm{eff}}$ vs. $\log T_{\mathrm{eff}}$ 
plot.We see that if we assign an evolutionary mass to a rotating star on the basis of a track 
without rotation, we obtain a mass which may be up to a factor of 2 too large. 
{\emph{Appropriate tracks have to be used for rotating stars}}.  The non--respect of
this prescription is likely the explanation of the so--called problem of the
mass discrepancy.
\section{Surface abundances}

\begin{figure}
\plottwo{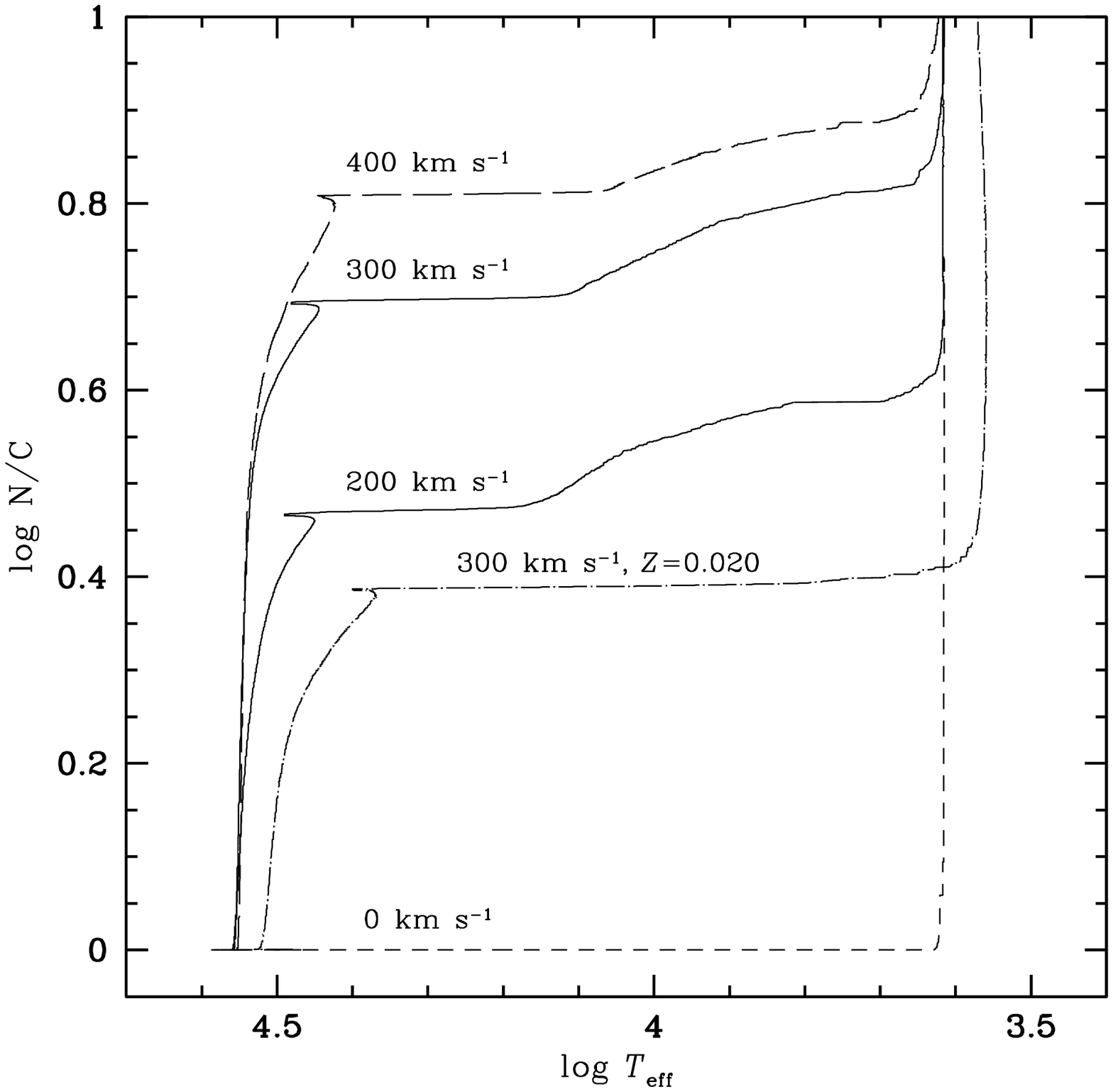}{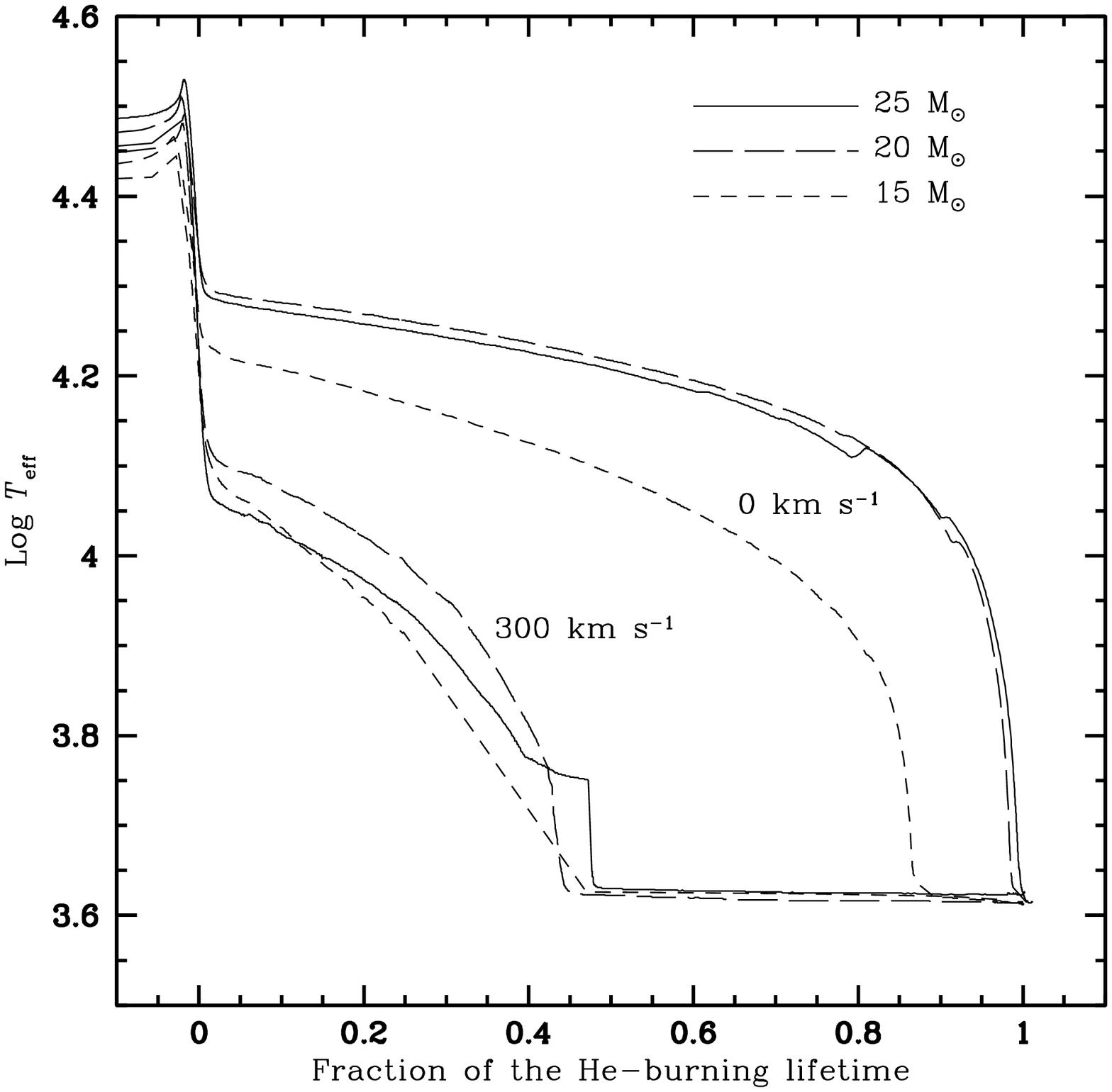}
\caption{Left: Evolution as a function of $log T_{\mathrm{eff}}$ of the abundance ratio N/C
(in number),
normalized to their initial values. The tracks are for $Z=0.004$. A model at Z=0.02
is also shown. The track with 0 km/s applies to both $Z=0.020$ and $Z=0.004$.
 Right: Evolution of  $T_{\mathrm{eff}}$  as a function of the fraction of lifetime
 spent in the He--burning phase fo 15, 20 and 25  M$_{\odot}$ at $Z=0.004$ for
 $v_{\mathrm{ini}}=0$ and 300 km/s, (Maeder \& Meynet 2001).} 
\end{figure}

As a result of internal mixing, He and N are brought to the surface, while C is
depleted. At solar $Z$,  the N/C ratios  are increased  by 
a factor 2--3 for a  20 M$_{\odot}$ model (Fig. 
7 left). For lower $Z$, we see that the enrichment in N/C is much larger.
This results from the larger shears in low metallicity models 
(Fig. 3). The higher N--enhancements at lower $Z$ are  consistent with
 the results by Venn (1998, 1999), who has found excesses of N/H up to about a factor of 10
 in supergiants of the SMC. At $Z=0.004$, the models show no sign of primary nitrogen.
Models at $Z=10^{-5}$ by Meynet \& Maeder (2002) show large amounts of primary nitrogen, which 
change the sum of CNO elements in advanced stages. 

There is still  a problem (Herrero et al. 2000). There are many
fast rotating stars, which have He excesses. The models indicate that
when the products of CNO burning appear at the surface, the  velocities
have already significantly decreased. A possible explanation is that
there are initially many very fast rotating stars, where the  enrichments appear
 quickly. Also, nuclear reactions in massive stars  are  active
in the pre--MS phase and thus  mixing may already occur during  the
pre--MS phase.

\section{The blue to red supergiant ratio}

In clusters of low $Z$, as in the SMC, there are lots of red supergiants.
However, the models of massive stars at low $Z$  do not reach in general the
red stage. If they do it, this happens only very late in evolution, so that there are
no red supergiants predicted. Fig. 7 (right) illustrates the point:  
for zero rotation most of the He--burning phase is spent in the blue and very little in the red.

Rotation very much changes the above results. Fig. 7 (right) shows that the models of 15 to 25 
M$_{\odot}$ with rotation spend at least  half of the He-burning lifetime in the red.
This is in agreement with the  blue to red supergiant ratios
of  0.5--0.8 observed in the SMC. The physical reason why rotation makes more red supergiants
is the following one: with rotation, the He--burning core is larger and there are 
also more He mixed just outside the core, in particular at the location of the H--burning shell.
Since the core is larger, and as there  is also more He outside, the H--shell is
 less active and the opacity is also lower. Thus, while there is an intermediate convective
 zone in non--rotating models, there is none in models with rotation. Convection means a 
 polytropic index $n=1.5$ and thus a high degree of compactness, which keeps the
 non--rotating stars in the blue. With rotation, the absence of intermediate convective zone
 permits the envelope to largely expand and the star moves toward the red supergiant stage.
 
 On the whole, we see that rotation is an essential ingredient of stellar models and that
 rotation modifies all the model outputs.\\
 
 {\emph{Acknowledgements:} We express our thanks to Raphael Hirschi for a careful reading 
 of the manuscript.}


\begin{references}
\reference Chaboyer, B., Zahn, J.P. 1992, A\&A 253, 173

\reference Herrero, A., Puls, J., Villamariz, M.R. 2000, A\&A 354, 193

\reference Kippenhahn, R., Thomas, H.C. 1970, in Stellar Rotation, IAU Coll. 4, Ed. A. Slettebak,
           Gordon and Breach, p. 20
\reference Lamers, H.J.G.L.M., Snow, T.P., Lindholm, D.M. 1995, ApJ 455, 269
\reference Maeder, A. 1999, A\&A 321, 134 (paper II)
 
\reference Maeder, A. 1999, A\&A 347, 185 (paper IV)

\reference Maeder, A. 2002, A\&A 392, 575 (paper IX)

\reference Maeder, A., Desjacques, V. 2000,  A\&A 372, L9

\reference Maeder, A., Grebel, E. \& Mermilliod, J.-C. 1999, A\&A 346, 459

\reference Maeder A., Meynet G. 2000,  A\&A 361, 159 (paper VI)

\reference Maeder A., Meynet G. 2001,  A\&A 373, 555 (paper VII)

\reference Maeder A., Peytremann, E. A\&A, 7, 120

\reference Maeder A., Zahn J.P. 1998, A\&A, 334, 1000 (paper III)

\reference Martin, E.L., Basri, G., Gallegos, J.E. et al. 1998, Ap.J. 499, L61

\reference Meynet G., Maeder A. 1997,  A\&A, 321, 465 (paper I)
 
\reference Meynet G., Maeder A. 2000,  A\&A, 3, 101 (paper V)

\reference Meynet G., Maeder A. 2002,  A\&A, 390, 561 (paper VIII)

\reference Talon, S., Zahn, J.P. 1997, A\&A 317, 749

\reference Venn, K.A. 1998, in Boulder--Munich II: Properties of Hot Luminous
         Stars, ed. I. Howarth, ASP Conf. Ser., 131, 177
         
\reference Venn, K.A. 1999, ApJ 518, 405 

\reference Vink, J.S., de Koter, A. \& Lamers, H.J.G.L.M. 2001, A\& A 362, 295

\reference Walborn, N. 1976, ApJ,  205, 419

\reference Zahn J.P. 1992, A\&A 265, 115


\end{references}
\end{document}